\documentclass[structabstract]{aa}
\usepackage{graphicx}
\usepackage{txfonts}

\newcommand{\TP}{\emph{$P$-$T$} profile}

\newcommand{\hd}{\object{HD~209458b}}

\begin{document}
   \title{TiO and VO broad band absorption features in the optical spectrum of the atmosphere of the hot-Jupiter \hd}
   \subtitle{}
   \titlerunning{TiO and VO broad band absorption features in the atmosphere of \hd} %
   \authorrunning{J.-M. D\'esert et al.}

  \author{J.-M. D\'esert\inst{1} \and A. Vidal-Madjar\inst{1} \and A. Lecavelier des Etangs\inst{1} \and D.
   Sing\inst{1} \and D. Ehrenreich\inst{2} \and G. H\'ebrard\inst{1} \and R. Ferlet\inst{1}}

   \institute{
   Institut d'Astrophysique de Paris, CNRS (UMR~7095) -- Universit\'e Pierre \& Marie
   Curie;
   98 bis, boulevard Arago 75014 Paris, France, \email{desert@iap.fr}
   \and
    Laboratoire d'astrophysique de l'observatoire de Grenoble, Universit\'e Joseph Fourier, CNRS (UMR 5571), BP53 38041 Grenoble cedex 9, France}

   \offprints{desert@iap.fr}

   \date{Received June xx, 2008; accepted September xx, 2008}


  \abstract
    {The presence of titanium oxide (TiO) and vanadium oxide (VO) gas phase species is searched for in the atmosphere of the hot Jupiter \hd.}
  {We compared a model for the planet's transmitted spectrum to
multi-wavelength eclipse-depth measurements (from 3\,000 to
10\,000~\AA), obtained by Sing et al. (2008a) using archived
HST-STIS time series spectra. We make use of these observations to
search for spectral signatures from extra absorbers in the planet
atmosphere between 6\,000 and 8\,000~\AA.}
   {Along with sodium depletion and Rayleigh scattering recently
published for this exoplanet atmosphere, an extra absorber of
uncertain origin, redward of the sodium lines, resides in the
atmosphere of the planet. Furthermore, this planet has a
stratosphere experiencing a thermal inversion caused by the
capture of optical stellar flux by absorbers that resides at
altitude. Recent models have predicted that the presence of TiO
and VO in the atmosphere of \hd\ may be responsible for this
temperature inversion. Although no specific TiO and VO spectral band head signatures have
been identified unambiguously in the observed spectrum, we suggest
here that the opacities of those molecules are possible candidates
to explain the remaining continuous broad band absorption observed
between 6\,200 and 8\,000~\AA. To match reasonably well the data, the
abundances of TiO and VO molecules are evaluated from ten to one
thousand times below solar. This upper limit result is in
agreement with expected variations with altitude due to depletion
effects such as condensation.}

   {}
  \maketitle

\section{Introduction}
\label{sec:intro}

The discovery of transiting extrasolar giant planets (EGP) has
opened the window to the direct detections and characterization of
their atmospheres. Because of the wavelength-dependent opacities
of absorbing species, measurement of relative changes in eclipse
depth as a function of wavelength during primary transit has the
potential to reveal the presence (or absence) of specific chemical
species (\cite{Seager2000a}; \cite{Hubbard01}; \cite{Brown01}). In
the case of \hd's atmosphere, the transmission spectroscopy method
led  to the detection of sodium (Charbonneau et~al. 2002). The
small atmospheric sodium signature has made its detection from
ground-based telescopes difficult. Only recently, Redfield et al.
(2008) and Snellen et al. (2008) reported the detection of
Na\,{\sc i} in the atmosphere of HD189733b and HD209458b,
respectively. Absorptions of several percents for H\,{\sc i}
Lyman-$\alpha$, O\,{\sc i} and C\,{\sc ii} have been measured in
the hydrodynamically escaping upper atmosphere (\cite{Madjar03},
2004, 2008). More recently, Rayleigh scattering by H$_2$ molecules
has been identified (Lecavelier et~al. 2008b) and a
temperature-pressure (TP) profile with inversion was derived (Sing
et~al. 2008b). This temperature inversion leads to high
temperature at both low and high pressure. Note that this
temperature bifurcation was very well predicted by atmospheric
models of strongly irradiated planets (\cite{Hubeny03}). In the
lower part of the atmosphere ($\sim$30 mbar), the temperature is
found to be in the range $1\,900 -2\,400$~K (Lecavelier et~al.
2008b), which corresponds to the M/L/T brown dwarf regime, as
expected for a hot Jupiter such as \hd\ (\cite{Kirkpatrick05}).

In the cool atmosphere of sub-stellar objects like brown dwarfs
and close-in EGPs, the strongest absorption features are expected
to be those of alkali metals like sodium and potassium
(\cite{Seager2000a}; \cite{Sudarsky00}; \cite{Burrows00}). This is
due to the condensation and the rain out of elements in the
atmosphere which clears the atmosphere of most of the metals and
keeps the less refractory alkali metals (\cite{Burrows99};
\cite{Lodders99}; \cite{Burrows02}; \cite{Sudarsky03}). Thus, in
contrary to the spectrum of HD189733b (Pont et al. 2007;
Lecavelier des Etangs et al. 2008a), the spectrum of \hd\ should
be dominated by absorption from sodium Na\,{\sc i} and potassium
K\,{\sc i}.

However, depending on the effective temperature, a large number of
diatomic and polyatomic molecules are predicted to be present
according to various models of brown dwarf and hot Jupiter
(\cite{Burrows99}; \cite{Lodders99}; \cite{Allard01};
\cite{Lodders02}; \cite{Hubeny03}; \cite{Burrows06}). Among those
molecules and at a temperature above 1\,800 K, titanium oxide
(TiO) and vanadium oxide (VO) in gas phase equilibrium are most
probably present with a high abundance in strongly irradiated
planet atmospheres (Seager et al. 1998; \cite{Hubeny03};
\cite{Fortney07}). Furthermore, the low albedo measurements of
\hd\ (\cite{Rowe06}) rule out most of the absorbents, but TiO and
VO. Recently, the interpretation of visible observations obtained
during the transit of this planet (Sing et al. 2008a), as well as
the interpretation of near infrared observations taken during its
secondary eclipse (Knutson et al. 2008) require that HD 209458b
experience an inversion in the atmospheric \TP\ and a have
stratosphere (Sing et al. 2008b; Burrows et al. 2007c; Burrows et
al. 2008). Such an inversion could be due to the absorption of the
incident visible light by TiO and VO molecules (Fortney et al.
2007). Ongoing models (Burrows et al. 2008) claim that this
inversion must be caused by the capture of the incident optical
stellar flux by a stratospheric absorber that could potentially be
TiO and VO. Similarly other studies (\cite{Hubeny03};
\cite{Fortney07}) have more particularly investigated the effects
of TiO and VO which absorb much of the incoming stellar radiation
high in the atmospheres of the hottest CEGPs.

In this work, we use {\it STIS} spectra obtained during planetary
transit at two spectral resolutions (low and medium) to search for
the direct evidence of the presence of TiO and VO in the planet's
atmosphere. Both datasets are combined to extend the measurements
over the entire optical regime to quantify other possible
absorbers appearing in the transmission spectrum. Rayleigh
scattering and sodium absorption have been proposed to explain the
spectral features observed between 3\,000~\AA\ and 6\,000~\AA\
(Lecavelier des Etangs et al. 2008b; Sing et al. 2008a; 2008b)
yet, additional absorptions remain unattributed from 6\,100~\AA\
up to 8\,000~\AA. These absorptions cannot be due to optically
thick high altitude clouds, which would otherwise mask the
detected signature of Rayleigh scattering (Lecavelier des Etangs
et al. 2008b). Since TiO and VO molecules are expected to be
present in high abundance for the range of temperatures and
pressures for strongly irradiated hot Jupiter, their opacities
could largely contribute to the observed absorption in this
wavelength domain (Seager et al.1998; \cite{Burrows99};
\cite{Sharp07}).

After a brief description of the observation and interpretation
status (Sect.~\ref{sec:obs}), we present, in
Sect.~\ref{sec:model}, the model together with the method used to
calculate the atomic and molecular opacities. Finally, we estimate
the contribution of  TiO and VO to the observed transmission
spectrum for several abundance scenarios (Sect.
~\ref{sec:analysis}) and discuss the consequences on the
atmospheric chemistry and aeronomy of HD209458b
(Sect.~\ref{sec:discuss}).


\section{Observations}
\label{sec:obs}

The \emph{HST-STIS} G750L, and G430L low resolution grating
observations of \hd\ analyzed here are also detailed in
\cite{Knutson07a}, \cite{Ballester07} and Sing et al. (2008a). For both
the G750L, and G430L gratings, two visits were done for each
grating, of five consecutive orbits each. Each visit span one
completed transit of \hd\ in 2003. Together they cover the
combined range 2\,900-10\,300~\AA, with some overlap around 5\,300~\AA\
with a resolving power $R=500$. While the broad-band
spectrophotometric analysis of \cite{Knutson07a} was used to
identify water absorption features in the atmosphere of \hd\ at
wavelength greater than 9\,000~\AA\ (\cite{Barman07}), here we
analyze the spectrum at full resolution.

We directly use \hd\ absorption depths (AD, see definition in
Sect.~\ref{sec:AD}) between 4\,000~\AA\ and 8\,000~\AA\ provided
by \cite{Sing08a}. The low resolution transit spectrum ratio, limb
darkening corrected, is plotted in Figure ~\ref{fig_atoms}. The
spectrum below 8\,000~\AA\ is considered here where the absorption
due to water molecules is negligible. Three prominent broadband
absorption features are revealed. The near ultraviolet (NUV)
absorption, at wavelength $\lambda$ $\le$ 5\,000~\AA, was first
reported by \cite{Ballester07} and explained by the absorption of
a hot hydrogen layer within the atmosphere. \cite{Lecavelier08b}
propose an alternative explanation invoking the Rayleigh
scattering by H$_2$ molecules. Within the same datasets,
\cite{Sing08b} found that the Na spectral line profile is
characterized by a wide absorption with a sharp transition to a
narrow absorption profile at higher altitudes values. This sharp
transition is interpreted by condensation or ionization which
deplete Na atoms in the upper atmosphere. Using a global fit to
these data, from 3\,000~\AA\ to 6\,200~\AA, \cite{Sing08b}
determine the average pressure-temperature profile ($P$-$T$) at
the planetary terminator. Two types of \TP\ were derived which
both can lead to the depletion of Na\,{\sc i} atoms. One profile
implies the condensation of Na\,{\sc i} and the other the
ionization of Na\,{\sc i}. We use this former one
(Fig.~\ref{fig_TP}). Both profiles experience hot temperatures,
above 1\,800~K at 50~mbar, due to Rayleigh scattering (Lecavelier
et al. 2008b). At such a temperature, TiO and VO are in gas phase.

A third broad absorption feature in the range 6\,200 - 8\,000~\AA\
appears on the AD spectrum and is still unexplained. This
absorption is composed of two remarkable features visible in
Fig.~\ref{fig_atoms}:

\begin{enumerate}

\item A strong and broad spectral absorption feature centered at
about 6\,250~\AA\ which is slightly smaller than for the Na\,{\sc i}
($AD=0.0148\%$ over ~100~\AA). Other spectral signatures appearing
at higher wavelengths are weaker than the 6\,250~\AA\ one.

\item A broad and continuous absorption level above 7\,000~\AA\
systematically higher than the lowest observational AD ($0.0146\%$
over 1500~\AA). This broad continuous flat absorption excess is
weaker than the spectral feature at 6\,250\AA\ but is equally as important
since it covers a large spectral domain, and has a much larger
absorption level than predicted by extrapolating Rayleigh
scattering and Na\,{\sc i} or K\,{\sc i} broad wings.

\end{enumerate}

We overplotted in Fig.~\ref{fig_atoms} the radii presented by
Knutson et al. (2008) that were derived using spectrophotometric
bandpass across roughly thousand angstrom bins. These radii are
found to be slightly different than those presented in this paper,
especially in the region of interest of this study (over
6\,000~\AA), though the differences are either close to or within
the error bars of the two studies. This differences could be
explained by the supplementary corrections in the data treatement
employed by Sing et al. (2008) who corrected the transmission
spectrum for telluric contamination and various systematics,
quantified the level of systematic ‘red noise’ to take it into
account in the computed errors and finally applied a spectral limb
darkening correction before binning. Note that the systematic
errors have a weak wavelength dependance, affecting the absorption
depth values over large wavelength regions in a similar manner,
largely preserving the wavelength structure of narrow bands such
as the 6\,250~\AA\ feature.

Looking at the low resolution spectrum, we find the minimum AD of
0.01453\% around 5\,000~\AA, which indicates a base pressure of
$33$ mbar in the \TP (Lecavelier des Etangs et al. 2008b).
Conversely, the maximum AD of 0.0149\% is obtained for the central
pixel of the Na\,{\sc i} line which corresponds to a pressure of
$1$ mb (Sing et al. 2008b).

Below we present possible identifications of absorbers that can
simultaneously explain the spectral features between 6\,200 and
7\,000 \AA\  and the broad relatively flat absorption excess from
7\,000 to 8\,000~\AA.


\section{Model description}
\label{sec:model}

In order to interpret the transit observations, we developed a
model that calculates the absorptions of atomic and molecular
lines in the atmosphere of a transiting planet.

Transit spectroscopy probes the transition region between the day
and night sides, called {\em limb}. We used the geometry of a
transiting system  (\cite{Brown01}) and a model originally
developed by Ehrenreich et al.\ (2006) and adapted for this work.
Along a single cord, stellar photons cross several levels of the
spherically stratified atmosphere of the planetary limb.

The total opacity, $\tau_{\lambda}$, along a cord (parallel to the
line of sight) as a function of its impact parameter $b$ and the
altitude $h$, is given as:

\begin{equation}
\label{eq:opacity} \tau_{\lambda,i}(b) = 2 \int_{0}^{+\infty}
\sigma_{\lambda,i} n_i(h) \mathrm{d} l,
\end{equation}

\noindent where $\sigma_{\lambda,i}$ is the cross section
coefficient for the species $i$ at the wavelength $\lambda$, and
$n_i(h)$ is the density of the species $i$ at an altitude $h$ in
the atmosphere (see Fig.~1 in Ehrenreich et al. 2006).


\subsection{Atomic and molecular line opacities}
\label{sec:opacities}

textbf{In the case of alkali atoms, we calculate the opacity using
the oscillator strength, the radiative lifetime and the
collisional broadening taken from Morton et al. (1991) and Iro et
al. (2005) which used Burrows et  al. (2000) for determining the
line shapes. Further progresses have been achieved in the theory
of these profiles (Burrows et al. 2003; Allard et al. 2005). The
difference between these new line shapes and the ones used in this
study resides mainly in the far wings. However, the contribution
of other absorbents dominates the spectrum in the far wings of the
alkali lines. Thus, we found that the calculations of Morton et
al. (1991) and Iro et al. (2005) were precise enough for our
study.

In the case of molecules, we calculate the line strengths for each
line of each species using the available data. We derive the
strength at local thermal equilibrium (LTE), noted $S$, of the
molecular spectral line from the general equation
(\cite{Sharp07}):

\begin{equation}
S = \frac{\pi e^2 g_if_{ij}}{m_ec} \frac{e^{-hcF_i/kT}}{Q(T)}
\left[1 - e^{-hc(F_j-F_i)/kT}\right]\, ,
\label{oscil}
\end{equation}

\noindent where $g_i$ and $f_{ij}$ are the statistical weight of
the $i^{th}$ energy level and the oscillator strength for a
transition from that level to a higher level $j$, respectively.
The excitation energies $F_i$, and $F_j$ are the term values in
cm$^{-1}$ of the $i^{th}$ and $j^{th}$ levels participating in the
transition. In addition, $Q(T)$ is the partition function of the
species at some temperature $T$. We compute the partition function
as described by \cite{Sauval84}, considering a fourth degree
polynomial function of $\log(\frac{5040}{T})$. The molecular lines
are broadened using a Voigt profile. At pressure $P$, the
collisional broadening is calculated using (Sharp \& Burrows
2007):

\begin{equation}
\Delta\nu_{Lorentz}=w_oP,
\label{dnu_lorentz}
\end{equation}
with $w_o=0.1$ cm$^{-1}$~bar$^{-1}$

Finally, we obtain the total monochromatic opacity of the
atmosphere by summing the individual contributions for each atom
and molecule weighted by their respective abundances at each
altitude level.

\subsection{Grid of opacities}
\label{sec:grid}

Computing the opacities directly at each altitude with the
corresponding temperature and pressure would require intense
calculation. Instead, we precalculate a grid of opacities for each
molecule. Our grid is composed of 42 cells with a range of
temperature from 100 to 5\,000~K and a range of pressure from 1 to
$10^{-6}$ bar. Opacities have been computed from 2\,900 to
10\,000~\AA\ with a resolution of 0.005~\AA\, which corresponds to
5 million elements per cell. For a given \TP\, we then calculate
the corresponding opacities by applying a linear interpolation on
the grid cells.

\subsection{Absorption Depth (AD)}
\label{sec:AD}

The output of the model is a spectrum ratio  defined by Brown
(2001) as the ratio of the flux received during the transit with
the flux received when the planet is not occulting the star of
radius $R_\star$. Otherwise indicated by the absorption depth
(AD), that is the occultation of the surface of the star by the
surface of the planet at each wavelength. The surface of the
planet is twofold: the optically thick disk of radius $R_P$ and
the wavelength-dependent occultation by the atmosphere that
surrounds the planetary disk expressed as an equivalent surface.
We used $R_{P}=1.32~R_{jup}$ and $R_{\star}=1.125~R_{\odot}$ from
\cite{Knutson07a}. \cite{Lecavelier08b} show that the observed AD
is well approximated by:

\begin{equation}
AD_{\lambda}=AD_0(1+\frac{2H}{R_P}\ln\frac{\sigma_{\lambda}}{\sigma_{\lambda_0}})
\label{ADbin}
\end{equation}

\noindent where AD$_0$ is AD at $\lambda=\lambda_0$ and H the
scale height. Thus the observed mean AD over a wavelength range of
a given spectral element (typically 50~\AA), is proportional to
the mean of the logarithm of the cross section. Therefore, we
calculate the effective cross section in a given spectral element
by averaging the logarithm of the cross section calculated at a
much larger resolution:

\begin{equation}
\sigma_{bin}[\lambda_i ; \lambda_j]=
exp \left( \int_{\lambda_i}^{\lambda_j}  \frac{\ln
\sigma_{\lambda}}{\lambda_i - \lambda_j}d_{\lambda}\right)
\end{equation}


\begin{table}
\caption{List of the most probable species with their
corresponding solar abundances which have significant absorptions
in the 6\,000-8\,000~\AA\ wavelength domain (from \cite{Lodders02} and
\cite{Sharp07}). This list is ordered by the product of an average
of the cross section between 4\,000 and 8\,000~\AA\ and its
corresponding solar abundance. This ordered list gives a rough
estimate of the plausible detectability for each atomic and
molecular species.}

\label{table:solar_abund}      
\centering                        
\begin{tabular}{c c c c}        
\hline\hline                 
Element & $\chi_{\odot}$ & $<\sigma_{\lambda}>$ ($cm^{2}$)& $\chi_{\odot}<\sigma_{\lambda}>$ ($cm^{2}$) \\    
\hline                        
   Na\,{\sc i}       & 3.e-6  &  1.e-19  & 1.e-25    \\
   K\,{\sc i}        & 2.e-7  &  1.e-20  & 1.e-27    \\
   Li\,{\sc i}       & 1.e-9  &  1.e-19  & 1.e-28    \\
\hline                        
   SiO       & 1.e-5   & 1.e-17  & 1.e-22    \\
   TiO       & 1.e-7   & 1.e-16  & 1.e-23    \\
   VO        & 1.e-8   & 1.e-16  & 1.e-24    \\
   MgH       & 1.e-9   & 1.e-17  & 1.e-26    \\
   H$_2$O      & 1.e-3   & 1.e-24  & 1.e-27    \\
   FeH       & 1.e-9   & 1.e-19  & 1.e-28    \\
   CaH       & 1.e-11 & 1.e-17  & 1.e-28    \\
   CrH       & 1.e-10  & 1.e-19  & 1.e-29    \\

\hline                                   
\end{tabular}
\end{table}


\section{Analysis}
\label{sec:analysis}

Spectral signatures, line profiles and abundances vary depending
on the temperature and pressure vertical profile in the
atmosphere. Here we use the vertical \TP\ presented in Figure
~\ref{fig_TP} and two sodium mixing ratios from \cite{Sing08b} as
well as the Rayleigh scattering derived by \cite{Lecavelier08b}.
The AD of 0.01453\%, corresponding to a pressure of 33 mbar, is
taken as the reference at 1.32 Jupiter radius (Knutson {et~al.}
2007). Using those hypothesis, we calculate the absorption
spectrum with only the abundances as free parameter. We vary the
abundances to find the best fit to the data using a $\chi^2$
minimization between the data and the model.

\subsection{Atomic line:}
\label{sec:atoms}

The most important absorptions by atomic lines are those of alkali
metals, Na\,{\sc i}, K\,{\sc i} and Li\,{\sc i} (see
Table~\ref{table:solar_abund}). This table provides the lines with
the strongest abundance weighted cross-section in the considered
wavelength range. None of this atomic lines, nor H$\alpha$, can
explain these observed spectral feature at 6\,250~\AA.

In the low resolution transit spectrum, we do not detect
significant absorption in the K\,{\sc i} line at 7\,698~\AA\ nor in
the Li\,{\sc i} doublet at 6\,708~\AA. These non detections exclude
the possibility of solar abundance over the whole atmosphere for
these two species (see Fig.~\ref{fig_atoms}). Assuming a constant
mixing ratio, we constrain the K\,{\sc i} and Li\,{\sc i}
abundances to be lower than $2\times 10^{-3}$ and $2\times
10^{-1}$ solar, respectively (1$\sigma$). Of course, these upper
limits does not exclude larger abundances below the altitude
corresponding to the observed absorption depth of $\sim$0.0145\%.

We conclude that the absorption depth measured between 6\,200 and
8\,000~\AA\ cannot be explained by the strong absorption lines
from abundant atomic species. The line profile of Na\,{\sc i} does
not affect the results on TiO/VO abundances since the remaining
broad band absorption begin at 200~\AA\ away from the center of
the line on the red side. In the following, we consider possible
absorptions by molecules.

\subsection{Molecular line:}
\label{sec:molecules}

A list of plausible molecular species absorbing in the 6\,000~\AA\
region is presented in Table~\ref{table:solar_abund}.

Although the molecules CH$_4$, NH$_3$ and CO are the most
important source of opacities in the infrared, their effects are
negligible in the visible wavelength domain; they are not
considered in the following analysis.

Using the abundance weighted mean cross-section list, we found
that most plausible species are SiO, TiO and VO respectively.
Silicium oxide (SiO) absorption mainly occurs below 2\,000~\AA,
outside our wavelength domain. As a result, TiO and VO are
predicted to be the most abundant and the most important sources
of opacity in the present domain (\cite{Fortney06};
\cite{Lodders02}). However, since both species condense, their
abundances decrease below 1500K.

Amongst all the other molecular absorbers that can be used to
explain the observation in this wavelength range, the hydride
metals CrH, FeH, MgH, and CaH were considered as potential
candidates to explain the shape of the spectrum. However, the
abundance of those elements drops with decreasing temperature
below 2\,000K, due to the formation of condensates
(\cite{Lodders06}; \cite{Kirkpatrick05}). Furthermore, their
absorption domain does not match the one considered here, like for
H$_2$O.

Among plausible molecules in the atmosphere of \hd, TiO and VO
have the largest cross-section in the visible domain. We explore
the extent to which various abundances of TiO and VO can be used
to reproduce both the medium and low resolution spectra. We
derived the LTE strength $S$ (Eq.~\ref{oscil}) of the molecular
spectral lines from \cite{Plez98}. These data list about 3 million
lines from 4\,000 to nearly 30,000 cm$^{-1}$, covering the whole
visible part of the spectrum for the transitions of the most
abundant isotopes $^{48}$TiO, together with $^{46}$TiO,
$^{47}$TiO, $^{49}$TiO, and $^{50}$TiO, for nine different
electronic band systems. Using the grid previously described in
Sect.~\ref{sec:grid}, we compute theoretical AD including TiO and
VO absorption at each wavelength (see Fig.~\ref{fig_sigma}). At a
temperature of 2\,000~K and a pressure of 0.1 bar, the lines are
significantly broadened, thus the rapid fluctuations in
cross-section over short wavelength intervals are suppressed
revealing the main band features. At significantly lower
pressures, the broadening of the lines is much smaller, the
cross-section shows rapid variations as a function of wavelength
and the broad main features do not show up so clearly.

In the following sections, we explore different scenarios for the
atmospheric abundance of TiO/VO in order to adjust our model to
the observations. We fit the data with an atmospheric model
including Na I, Rayleigh scattering, and TiO and/or VO in
different abundances. In a first step, we will consider a unique
abundance of TiO/VO over the whole atmosphere. In a second step,
we will assume two levels of abundance of TiO/VO. We finally
discuss the fits obtained using those different abundances.

\subsection{TiO  :}
\label{sec:TiO}

As a first step, we consider only TiO and constrain the abundance
by fitting the  observational low resolution spectrum ratio.
Starting with the simplest assumption, we consider a solar
abundance over the whole atmosphere of the planet.

Our model gives an AD average of 0.155\% over the full bandpass
4\,000-8\,000~\AA\ (see upper dashed-dotted line in
Fig.~\ref{fig_molecules}a). Such a value is significantly above
what is observed. Therefore, TiO abundance must be much lower than
solar above altitudes corresponding to the observed AD level.

Assuming a constant mixing ratio, the best fit of the low
resolution data set over the region 4\,000-5\,700~\AA\ gives a TiO
abundance of $(8.\pm0.6)\times 10^{-4}$ solar. With this TiO
abundance, the $\chi^2$ is reduced from 1754 to 1207 for $n = 942$
degrees of freedom (using non rebinned spectra). However, this
abundance cannot reproduce the 6\,250~\AA\ spectral feature and the
flat absorption of the observations between 6\,200-8\,000~\AA. We
conclude that other absorbers must be considered to explain this
part of the AD spectral profile.

\subsection{TiO and VO:}
\label{sec:TiO_VO} Since VO is the most important source of
opacity after TiO (see Table~\ref{table:solar_abund}), we add this
molecule to our model. The chemistry of the vanadium is quite
similar to the one of titanium, where the monoxide is in the gas
phase at a temperature higher than the temperature of
condensation.

In a first step, we consider a solar abundance over the whole
atmosphere. As for TiO, VO with solar abundance cannot reproduce
the observations, especially around 5\,000~\AA. The abundance of
VO must be lower. Assuming constant mixing ratios for both
molecules, the best fit gives for TiO $(6\pm0.6)\times 10^{-4}$
and for VO $(3\pm0.5)\times 10^{-2}$ as shown in
Fig.~\ref{fig_molecules}b. We obtain a satisfactory fit with a
$\chi^2$ of 817, or a $\chi^2/n \approx 0.86$ in the low
resolution data set. The continuous flat part of the AD curve can
be well reproduce by absorption of VO molecules. The region around
5\,000~\AA\, already well reproduced with Rayleigh scattering and
with sodium lines absorptions, is not affected by the presence of
TiO and VO molecules. However, the spectral features around
6\,250~\AA\ remains unexplained assuming constant mixing ratios
for both species. Note that sligthly lower abundances are
necessary for TiO and VO if we consider the Knutson et al.(2007)
results for the radii over 6\,000\AA.

\subsection{TiO, VO, and condensation:}
\label{sec:TiO_VO_cond}

With a large temperature gradient, TiO and VO abundances are
expected to vary along the vertical atmospheric profile.
Furthermore, the \TP\ derived by \cite{Sing08b} crosses the TiO
and VO condensation curve (See Fig.~\ref{fig_TP}). To check if
signature of condensation can be found in the present data set, we
assume in this section that TiO/VO molecules can be depleted above
a given altitude. Indeed, TiO and VO are expected to condense
into many Ti and V-bearing compounds as described in details by
Lodders (2002). To implement this assumption, we introduce a
maximum altitude, corresponding to an absorption depth of
condensation ($AD_{cond}$), above which the absorption by the
corresponding molecules in gas phase vanishes.

In the model, we derive the altitude of condensation from the
intersection between the \TP\ and the condensation curve. The
condensation curves of TiO and VO depend mainly on the
temperature (see Fig.~\ref{fig_TP} and \cite{Sharp07}). The TiO
condensation point occurs at a temperature of 1\,850~K and a
pressure of 0.02 bar (see Fig.~\ref{fig_TP}). This altitude
corresponds to an AD of condensation $AD_{cond}=0.0146\%$. In the
case of VO, the condensation point is at 1\,600~K and 0.01 bar,
leading to $AD_{cond}=0.0147\%$ (see condensation limits in
Fig.~\ref{fig_molecules}c).

Using the \TP\ presented in Fig.~\ref{fig_TP}, TiO and VO
$AD_{cond}$ levels are lower than the 6\,250~\AA\ spectral feature
AD peak. However, both $AD_{cond}$ levels are found to be above
the observed average AD, above the continuous flat part between
6\,500~\AA\ $-$ 7\,500~\AA, and significantly above the lowest
$AD=0.0144\%$ around the left wing of the Na\,{\sc i} at 5\,000~\AA.
Thus, these molecules cannot be in solar abundances below the
altitude of condensation, otherwise the resulting AD curve would
be at the level of the condensation limits. We fit the AD curve
between 4\,000-8\,000~\AA\ and obtain an abundance for TiO and VO
below their altitudes of condensation of $(8\pm0.5)\times 10^{-4}$
and $(6\pm0.4)\times 10^{-2}$ solar, respectively. We obtain a fit
with a similar $\chi^2$ as before (Sect.~\ref{sec:TiO_VO}) with
one level of abundance for TiO/VO over the whole atmosphere.

Although the "cold-trap" can explain the lowest observational ADs,
our assumption of null abundances above this altitude cannot
describe the observed 6\,250~\AA\ spectral signature. Thus TiO/VO
molecules should remain in the upper part of the atmosphere if the
feature is due to these species.

\subsection{TiO and VO in two separated layers:}
\label{sec:TiO_VO_cond2}

In this section, we explore the possibility of the presence of TiO
and VO molecules in the upper part of the atmosphere above the
condensations altitudes. In the lower part of the atmosphere, we
keep the abundances found in Sect.~\ref{sec:TiO_VO_cond}. Then a
level of null abundance is imposed, corresponding to the
condensation level. Finally another abundance level is chosen for
the upper part of the atmosphere, where the temperature rise up
and cross again the condensation curves of TiO/VO. In the upper
part of the atmosphere, where the pressure is lower, the TiO lines
are less pressure broadened, numerous narrow peaks appear in the
TiO opacity curve, in particular, a strong peak appears at 6\,200
\AA\ (Fig.~\ref{fig_molecules}d).

The best fit gives a TiO and VO respectively $(1\pm0.2)\times
10^{-4}$ and $(1\pm0.2)\times 10^{-2}$ solar abundances in the
upper layer, above the altitudes of condensation. However the
$\chi^2$ does not significantly decrease (802 for 944 degrees of
freedom). The spectral features of the AD curve is now better
fitted with TiO, at least to first order, whereas VO molecules
reproduce mainly the continuous flat part. In that way, We can
reproduce more successfully the spectral feature, the continuous
flat part and match the Na\,{\sc i} blue wing simultaneously. Note
that the flat part of the spectrum ratio cannot be reproduced if
we choose a null abundance of TiO/VO in the lower part of the
atmosphere.

\begin{table}
\caption{Bests $\chi_{2}$ fits performed for various model and the derived TiO and VO abundances.}

\label{table:Chi2}      
\centering                        
\begin{tabular}{c c c c}        
\hline\hline                 
Model & $\chi_{2}$ & TiO $(\times 10^{-4}$ & V0 $(\times 10^{-2})$  \\    
\hline                        
   Only TiO                                           & 1207  &   $(8.\pm0.6)$  &  $0$  \\
   TiO and VO                                     & 817   &   $(6.\pm0.6)$  &  $(3.\pm0.5)$    \\
   TiO and VO with condensation       & 816   &   $(8.\pm0.5)$  &  $(6.\pm0.4)$    \\
   2 levels of TiO and VO                   & 802   &   $(1.\pm0.2)$  &  $(1.\pm0.2)$    \\

\hline                                   
\end{tabular}
\end{table}


\section{Discussion}
\label{sec:discuss}

We have highlighted that the \hd\ atmosphere is optically thick at
low pressures and this requires new absorbers between
6\,000-8\,000~\AA. The observation of the Na\,{\sc i} in the
atmosphere exclude the presence of clouds at pressure under $30$
mbar in the observed limb of the planet. We have shown that
Li\,{\sc i} and K\,{\sc i} absorption cannot reproduce the broad
band absorptions observed. Nonetheless, we derive upper limits for
their abundances at $2\times 10^{-3}$ solar for K\,{\sc i} and
$2\times 10^{-1}$ for Li\,{\sc i}.

The Barman (2007) analysis of the Knuston et al. (2007) results
and Sing et al. (2008b) analysis both agree that remaining
absorbents are required to explain the continuous broad band
absorption observed between 6\,200 and 8\,000~\AA, since the radii
derived in both reductions are significantly above the model with
no TiO nor VO molecules (see Fig.~\ref{fig_atoms}). As a
consequence, the presence of remaining absorbents, possibly TiO
and VO, is mandatory to explain the observations. This result does
not depend on the \TP\ adopted. The abundances of TiO and VO are
only poorly constrained by any \TP. We decided to use the \TP\
derived by Sing et al. (2008) as an example.  In that case, the
lower part of the \TP, i.e below the condensation curves, is the
most critical part for the determination of the TiO and VO
abundances. The abundances of TiO and VO molecules are mainly
constrained by the level of the observed AD curve between 6\,500
and 8\,000~\AA\ and by the pressure of the lower point. Indeed,
for this lower point, the TiO and VO lines are mostly pressure
broadened. Thus, the abundances of these species are mainly
sensitive to the pressure and not to the temperature in the lower
part of this profile. The \TP\ above the condensation curves
constrains mostly the spectral features, especially the
6\,250~\AA\ one, as seen for the case two level of TiO and VO.
However, we do not detect any TiO/VO narrow spectrale features,
thus any \TP\ above the condensation curve can be adopted. With
the \TP\ used in this study, we have shown that no clear solutions
emerge wether condensation is taken into account, or if two levels
of TiO and VO molecules are considered (see
Table~\ref{table:Chi2}). Thus, other profiles could led to the
same conclusion for the presence of TiO and VO and/or of potential
additional absorbers.

Although no typical TiO and VO spectral signatures have been
identified unambiguously in the observed spectrum, we suggest that
the opacities of those molecules are the best candidates to
explain the remaining continuous broad band absorption observed
between 6\,200 and 8\,000~\AA. Theoretical absorption with models
including TiO/VO with abundances below solar were evaluated to
match the data reasonably well. Using the \TP\ from
\cite{Sing08b}, we derived upper limits for the TiO and VO
abundances. The model without TiO or with TiO but no VO molecules
give the worst $\chi_{2}$ solutions. We found that the abundance
of TiO should be around $10^{-4}$ to $10^{-3}$ solar, and  the
abundance  VO around $10^{-3}$ to $10^{-2}$ solar. The fits become
marginally worse on the blue side of the sodium wing when adding
TiO and VO (see Fig.~\ref{fig_molecules}). This is because the
\TP\ used in our work was obtained by fitting the observation on
the blue side of the sodium line with a model which contains only
Rayleigh scattering and sodium opactity (Sing et al. 2008b) which
helped limit the number of free parameters in the fit, such that a
convergent solution could be found easilly. This model fits
precisely the observed data for this part of the spectrum (see the
dotted line in Figs.~\ref{fig_atoms} and ~\ref{fig_molecules}).
The addition of the TiO and VO opacities in this model,
contributes to slightly decrease the quality of the fit on the
blue side of the sodium line (see continous line in
Fig.~\ref{fig_molecules}). This is, however, of marginal
consequence on the TiO and VO evaluations in our study since these
are mostly controlled by the main part of the observations in the
redder range of the spectrum. Considering the quality of the
observations this does not affect the TiO and VO evaluations.
Hence, the use of the \TP\ with only Rayleigh and sodium
calculated by Sing et al.2008 is sufficient to derive reasonably
good estimates of the TiO and VO abundances.

Two layers of TiO/VO in the atmosphere are better suited to
explain the 6\,250~\AA\ spectral feature, however the $\chi^2$ is
not significantly improved (Table~\ref{table:Chi2}). Note that,
the 6\,250~\AA\ spectral feature is seen in the two independent
spectra obtained at both low resolution, as mentioned here, and at
medium resolution as observed by \cite{Sing08a} in the G750M STIS
spectra.

The tholins, polyacetylenes, or various non-equilibrium compounds
at high-altitude  could also be responsible for the  temperature
inversion. However, the study of these molecules requires a full
non-equilibrium chemical model which is not the purpose of that
work. TiO and VO molecules are very good high-altitude absorber
candidates (\cite{Hubeny03}; \cite{Fortney07}). Fortney et al.\
(2008) highlighted the importance of gaseous TiO and VO opacity in
their model of highly irradiated close-in giant planets. They
define two classes of irradiated atmospheres. Those which are warm
enough to have a strong opacity due to TiO and VO gases (``pM
Class'' planets), and those that are cooler (``pL Class'' planets)
dominated by Na\,{\sc i} and K\,{\sc i}. Our possible detection of
TiO/VO and Na\,{\sc i} in the hot atmosphere of \hd\ confirm that
this planet is located in the transition region between the two
classes defined by these authors. As observed here, due to the
presence of TiO/VO, \hd's atmosphere absorbs incident energy
between 6\,500~\AA\ and 8\,000~\AA. Consequently, this planet has a
hot stratosphere (around 2\,500~K) with a temperature inversion
(\cite{Burrows07c}). As another consequence, because of the
thermal emission of the energy trapped by the TiO/VO absorption,
the planet appears very bright during the mid infrared secondary
eclipse (\cite{Knutson08}).

The first titanium condensates appearing at high temperature and
high pressure are Ti$_3$O$_5$ and CaTiO$_3$ (\cite{Burrows99};
\cite{Lodders02}). The condensation of vanadium starts at lower
temperature (1\,600 K) than for titanium (1\,800 K). The vanadium
condenses into solid VO and then into V$_2$O$_3$. Thus, depletion
of TiO should start at lower altitude than for VO. Below 1\,600K,
TiO, VO and major refractory elements are absent, leaving
monoatomic Na\,{\sc i} and K\,{\sc i} to dominate the spectrum.
Nevertheless, sodium and potassium chlorides become increasingly
abundant with decreasing temperature, especially for KCl which is
the dominant K-bearing compound. Condensed potassium depletes the
atmosphere of atomic K\,{\sc i}, as seen with Na\,{\sc i}, leading
to reduced signatures. This could explain why Na\,{\sc i} is seen
in abundance in the low resolution data, but not K\,{\sc i}. In
the lower warmer atmosphere, where wide atomic K line wings would
be observable even at low resolution, TiO and VO likely make up
the surrounding continuum further masking the signature.

An alternative \TP\ with Na ionization has been also proposed
(Sing et al. 2008b). A model with Na ionization leaves a wide
range of temperatures possible in the middle atmosphere. Within
the framework of equilibrium chemistry, this profile presents a
lower temperature gradient \textbf{(Fortney et al. 2003)}. The
minimum temperature is below the TiO/VO condensation curves, and
above the sodium condensation curve. We did not enter into a
detailed study of TiO/VO abundances with these \TP; however, we
note that these profiles would have the effect to increase the
abundance of TiO and VO molecules in the middle atmosphere. See
Sing et al. (2008b) for more details about ionization.

If present in the upper part of the atmosphere, TiO/VO molecules
are not fully depleted by condensation. Thermochemical equilibrium
calculations with rainout (\cite{Burrows99}, \cite{Lodders02},
\cite{Hubeny03}) have shown that TiO and VO can exist at
high-$T$/low-$P$ points such as in the upper part of our \TP. The
presence of TiO and VO at high-$T$/high-$P$ points leads to a
situation in which there is two levels of TiO/VO. This has been
already proposed by Hubeny et~al. (2003) to explain the presence
of a temperature inversion in strongly irradiated planets
atmosphere's. Since the \TP\ used here crosses twice the Ti and V
condensation curves, two levels containing TiO/VO molecules in gas
phase could be separated by a middle level that is free of those
molecules, as proposed by Hubeny et al. 2003. Nonetheless, a cold
trap region is usually expected to deplete the upper low-$P$
region. Flushing out the upper atmosphere of TiO/VO would rule out
6\,250~\AA\ feature as being due to TiO. However, for now, it has
not been proven that cold-trap effect would deplete those
molecules, especially for the hydrodynamically escaping type of
upper atmosphere (\cite{Madjar03}, 2004, 2008; Lecavelier 2004;
Lecavelier et al. 2007; \cite{Ehrenreich08}).

New observations with higher signal-to-noise ratio and better resolution,
together with improved chemical models, are required to address
TiO and VO spectral absorption features and detect specific band
heads.


\newpage
\begin{figure*}
\includegraphics[width=12cm]{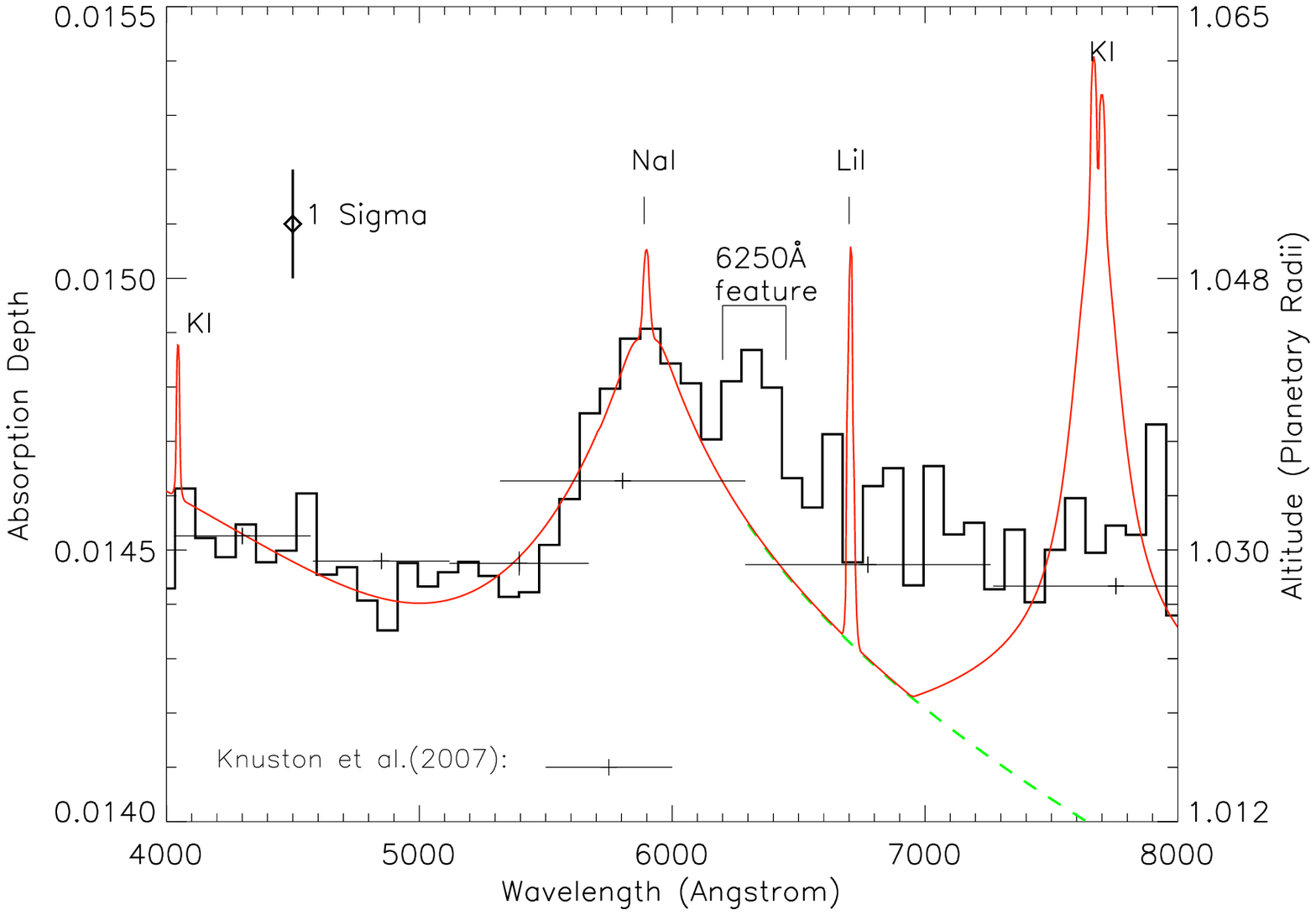}

\caption{The low resolution STIS measurements of the planetary
transit absorption depth (AD) corrected from limb-darkenning
effects and binned by 60 pixels (histogram). The observed
1$\sigma$ errorbar is plotted above the spectrum. The radii
derived by Knutson et al.(2007) with the corresponding error bars
and spectrale bins are plotted (squares). The dashed line
corresponds to the best fit model assuming Rayleigh scattering and
sodium absorption with the physical T-P profile plotted in
Fig.~\ref{fig_TP}. The difference between the dashed curve and the
observed AD spectrum for wavelength over 6\,000~\AA\ indicates
that remaning absorbents should be presents in this spectral
domain. Overplotted in continuous line, is the same model, with
atomic lithium and potassium with solar abundances.This model
cannot reproduce the observations above 6\,200~\AA.}
\label{fig_atoms}
\end{figure*}
\newpage

\newpage
\begin{figure*}
\centering
\includegraphics[angle=90,width=10cm]{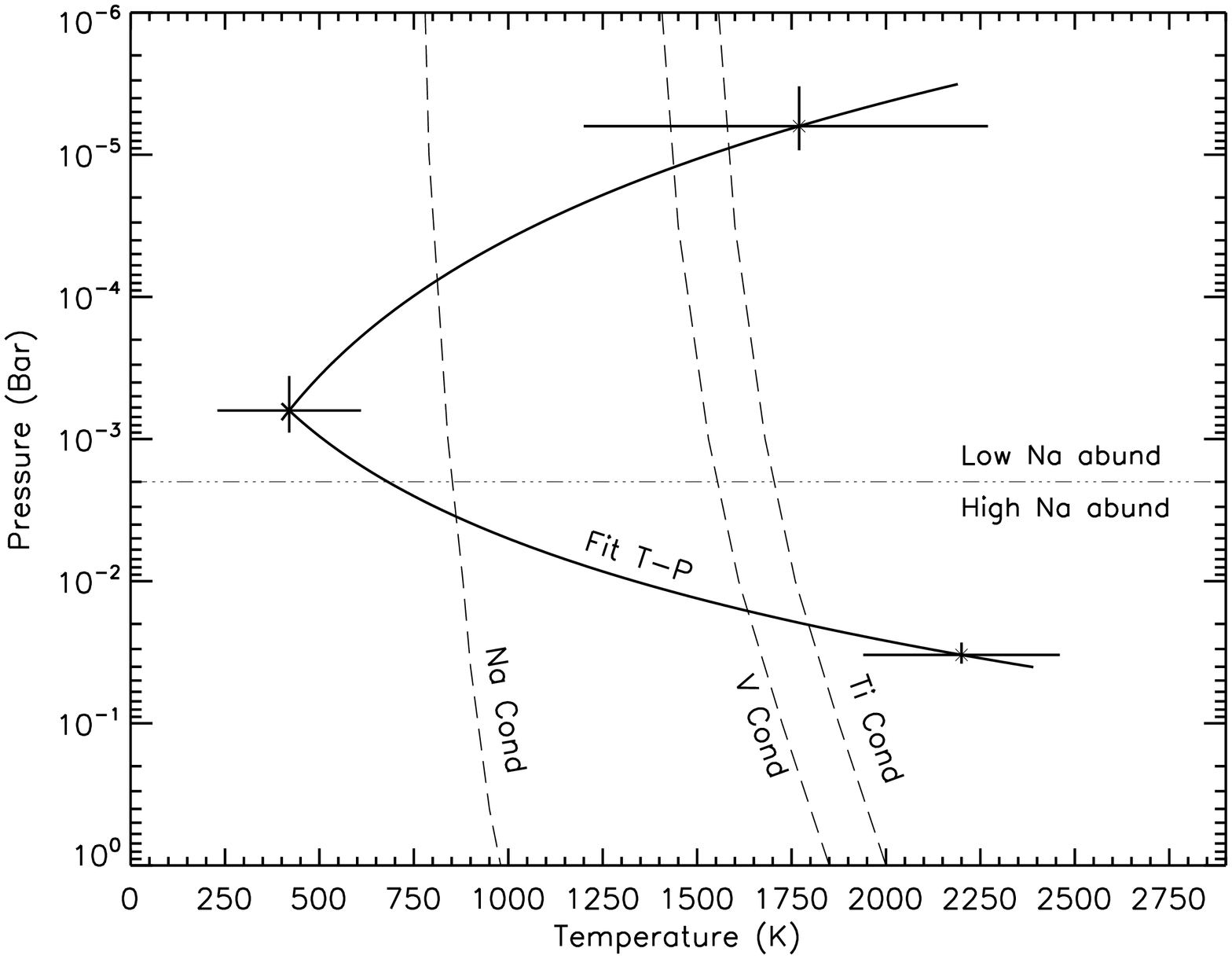}
\caption{The atmospheric Temperature-Pressure profile (\TP) with
error bars derived by \cite{Sing08b} used in this study. The
dashed lines correspond to the condensation curves for titanium,
vanadium and sodium. The three points of this T-P profile are
derived from the fit of the observed absorption depth curve. The
hot point at a pressure of 0.05~Bar is imposed by the Rayleigh
scattering (See Lecavelier et al. 2008).}
\label{fig_TP}%
\end{figure*}

\newpage
\begin{figure*}
\centering
\includegraphics[angle=90,width=10cm]{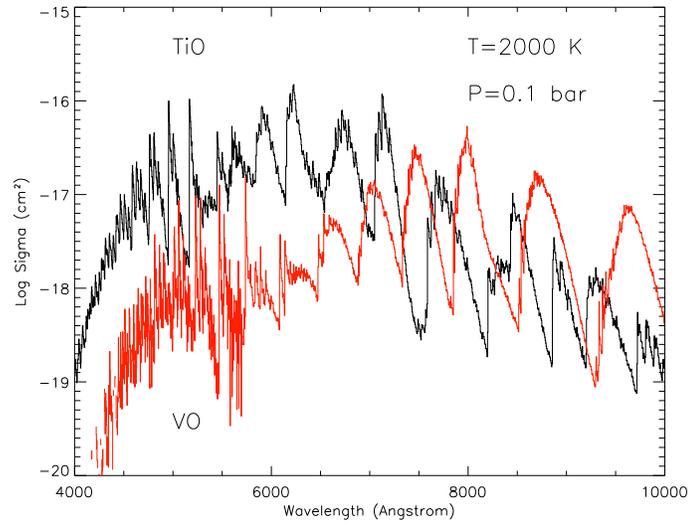}

\caption{The $\log_{10}$ of the monochromatic cross-section
$\sigma$ (cm$^2$) as a function of wavelength for the
vibration-rotation transitions of TiO and VO. The contribution due
to different isotopes is included. TiO has a
strong absorption feature shortward of 7\,500~\AA, and has a strong
peak near of 6\,200~\AA\ as observed in Fig.~\ref{fig_atoms}.}
\label{fig_sigma}%
\end{figure*}

\newpage
\begin{figure*}
\includegraphics[width=17cm]{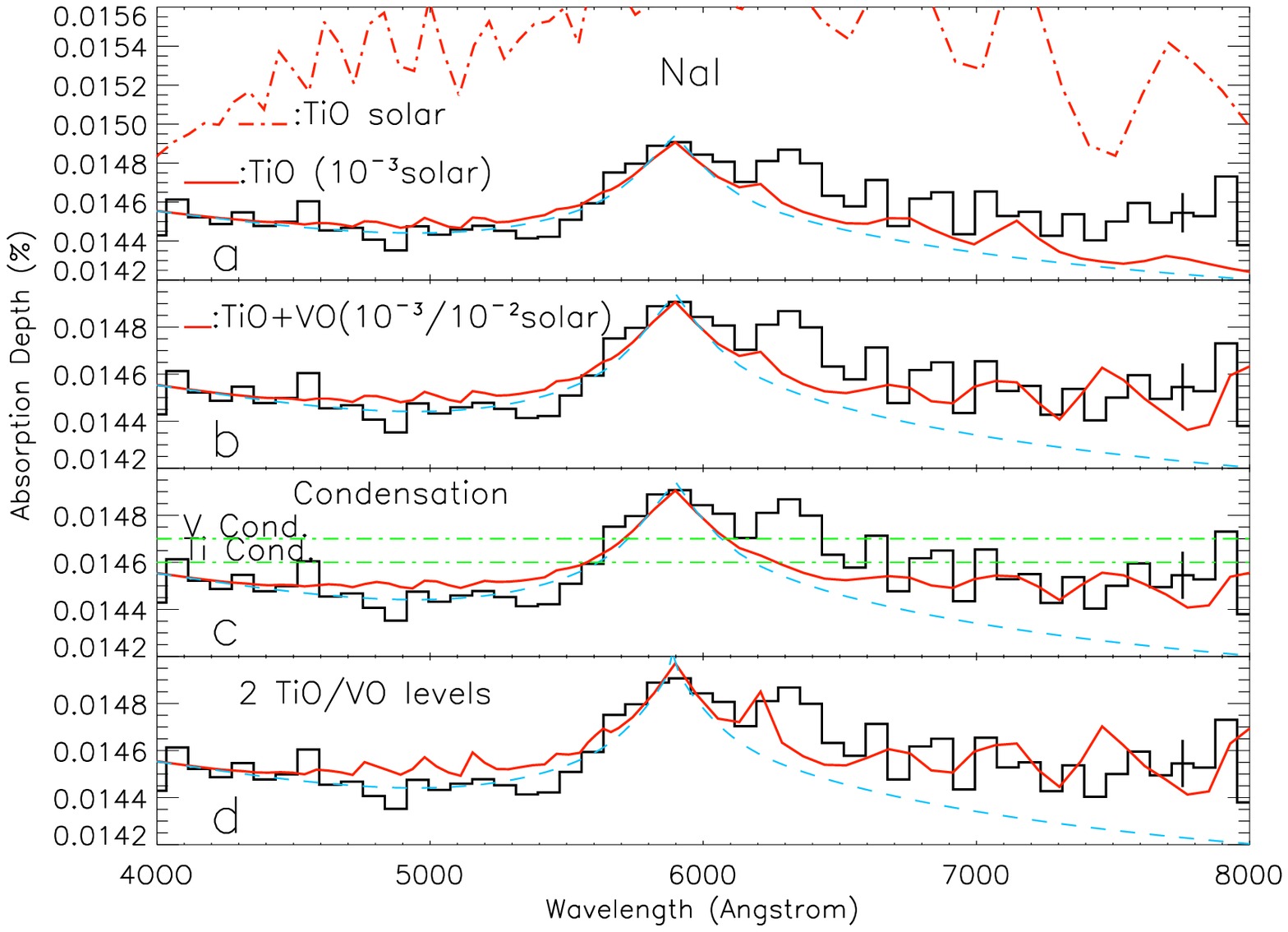}
\newline
\caption{Same as Fig.~\ref{fig_atoms} with fits obtain with
various models (continuous thick line). The observed 1$\sigma$
error bars are plotted around 7\,800~\AA\ (see also Fig.10 of Sing
et al. 2008). The dashed line shows the fit obtained using only
the model with Rayleigh scattering and sodium absorption with the
T-P profile (Fig.~\ref{fig_TP}). A remaining  broad band
absorption appears in between 6\,200 to 8\,000~\AA\ which could be
attributed to TiO and VO opacities. a: Model with a constant
mixing ratio of TiO. A constant solar abundance of TiO is excluded
(upper dashed-dotted line). The best fit gives an abundance of
$(8\pm0.6)\times 10^{-4}$ solar (red solid line). b: Model with a
constant mixing ratio of TiO and VO. Solar abundance for VO is
also excluded. The red solid line is the best fit which includes
TiO and VO. c: TiO, VO, and condensation. The altitudes of
condensation for titanium and for vanadium are both plotted (dash
dotted lines) together with the best fit (red solid line). d: Best
fit for two distinct levels containing TiO and VO (See
Sect.~\ref{sec:TiO_VO_cond2}). No clear solution emerge in between
the three last models (red solid line).} \label{fig_molecules}
\end{figure*}
\newpage



\begin{acknowledgements}
We thank our anonymous referee and our editor for their comments
that strengthen the presentation of our results. We also thank
Adam Burrows, Jonathan Fortney and Mark Marley for useful
discussions.
\end{acknowledgements}

\end{document}